\documentclass{article}%
\usepackage{amsmath}
\usepackage{graphicx}%
\usepackage{amsfonts}%
\usepackage{amssymb}
\begin{document}

\begin{center}
{\Large Perturbative Analysis of Wave Interaction in Nonlinear Systems}

\textbf{Alex Veksler}$^{1}$\textbf{ and Yair Zarmi}$^{1,2}$

Ben-Gurion University of the Negev, Israel

$^{1}$Department of Physics, Beer-Sheva, 84105

$^{2}$Department of Solar Energy \& Environmental Physics

Jacob Blaustein Institute for Desert Research, Sede-Boqer Campus, 84990
\end{center}

{\Large Abstract}

\begin{quotation}
This work proposes a new way for handling obstacles to asymptotic
integrability in perturbed nonlinear PDE's within the method of Normal Forms
(NF) for the case of multi-wave solutions. Instead of including the whole
obstacle in the NF, only its resonant part (if one exists) is included in the
NF, and the remainder is assigned to the homological equation. This leaves the
NF integrable and its solutions retain the character of the solutions of the
unperturbed equation.

We exploit the freedom in the expansion to construct canonical obstacles which
are confined to the interaction region of the waves. \ For soliton solutions
(e.g., in the KdV equation), the interaction region is a finite domain around
the origin; the canonical obstacles then do not generate secular terms in the
homological equation. When the interaction region is infinite (or
semi-infinite, e.g. in wave-front solutions of the Burgers equation), the
obstacles may contain resonant terms.

The obstacles generate waves of a new type which cannot be written as
functionals of the solutions of the NF. When the obstacle contributes a
resonant term to the NF, this leads to a non-standard update of the wave velocity.
\end{quotation}

{\Large Keywords: }nonlinear evolution equations, wave interactions, obstacles
to asymptotic integrability, perturbed KdV equation, perturbed Burgers equation.

\section{The problem of obstacles to asymptotic integrability}

The analysis of the effect of a perturbation on wave solutions of evolution
PDE's has evolved in two different approaches. In scattering approach, a
solution of the unperturbed equation is scattered off a perturbation that is
turned on at $t=0$. As the unperturbed solution is not a solution of the
perturbed equation, its amplitude decays and its wave number, velocity and
phase shift are modified. In addition, a soliton tail has been found outside
the soliton sector in the case of the perturbed KdV equation. The methods used
have been a combination of the theory of inverse scattering and the multiple
time expansion procedure \cite{1}-\cite{8}.

The second approach deals with a different issue. Integrable nonlinear
evolution equations are lowest-order approximations for the more complicated
equations of the full dynamical systems (e.g., the equations of Fluid Dynamics
in the cases of the Burgers \& KdV equations, and Maxwell's equations in the
case of the NLS equation). To improve the approximation, one includes
higher-order effects of the original physical system. In this case the
perturbation is not turned on at $t=0$. It exists for all times, going back
to$\;t=-\infty$.

In this approach, one is searching for a zero-order solution, which has the
same wave structure as the solution of the unperturbed equation, except for an
update of the wave velocity by the higher-order effects. The method of Normal
Forms was used for the analysis of solition solutions of the perturbed KdV and
NLS equations \cite{15}-\cite{17}, \cite{20}, \cite{25} and front solutions of
the perturbed Burgers equation \cite{14}, \cite{18}, \cite{23}. The method of
Multiple Time Scales was used in the case of the perturbed KdV equation
\cite{21}, \cite{22}, \cite{24} and in the case of the perturbed NLS equation
\cite{10}.

We focus on problems that arise in the analysis of wave solutions within the
second approach. Most integrable nonlinear evolution PDE's allow for single-
as well as multi-wave solutions. The multi-wave solutions usually asymptote to
well-separated single waves in the $x-t$ plane, except for interaction
regions, where the multi-wave character of the solution is lost. The
interaction regions may be localized (e.g., in the case of KdV-multi-soliton
solutions) or semi-infinite (e.g., Burgers-multi-fronts). The main purpose of
this work is to investigate the effect of a perturbation that is added to the
unperturbed equation on the wave solutions of nonlinear systems.

The perturbed equations are often analyzed by the method of Normal Forms (NF)
\cite{12} - \cite{14}, briefly described in the following. Let%

\begin{equation}
w_{t}=F^{(0)}[w]+\sum_{k=1}\epsilon^{k}\,F^{(k)}[w] \label{eq 1}%
\end{equation}

be a perturbed nonlinear evolution PDE (square brackets imply that the
corresponding term is a differential polynomial in $w\left(  x,t\right)  $).
We assume that $w\left(  x,t\right)  $ may be expanded in a power series in
the small parameter $\epsilon$ of differential polynomials of $u\left(
x,t\right)  $ (NIT - Near-Identity Transformation):%

\begin{equation}
w=\sum_{k=0}\epsilon^{k}\,u^{(k)}[u]\ \ \ \ \ \ (u\equiv u^{(0)}) \label{eq 2}%
\end{equation}

The time evolution of the zero-order term, $u(x,t)$, is assumed to be governed
by the Normal Form (NF):%

\begin{equation}
u_{t}=\sum_{k=0}\epsilon^{k}\,\alpha_{k}\,S^{(k)}[u],\;\;\;\;\left(
a_{0}=1\right)  \label{eq 3}%
\end{equation}

Here, $S^{(k)}$ are\ the \emph{resonant terms} usually called \emph{the
symmetries}. Their time dynamics is equivalent to that of $u\left(
x,t\right)  $ up to first order, that is,%

\begin{equation}
\left(  u+\mu\,\,S^{\left(  n\right)  }\right)  _{t}=F^{\left(  0\right)
}\left[  u+\mu\,\,S^{\left(  0\right)  }\right]
,\,\,\,\,\,\,\,\,\,\,\,\,\,\left(  \mu\ll1\right)  \label{eq 4}%
\end{equation}

As a result, their \emph{Lie Brackets} vanish:%

\begin{equation}
\left[  F^{\left(  0\right)  },S^{\left(  n\right)  }\right]  \equiv\sum
_{i}\left\{  \frac{\partial\,F^{\left(  0\right)  }}{\partial\,u_{i}}%
\partial_{x}^{i}\,S^{\left(  n\right)  }-\frac{\partial\,S^{\left(  n\right)
}}{\partial\,u_{i}}\partial_{x}^{i}\,F^{\left(  0\right)  }\right\}  =0
\label{eq 5}%
\end{equation}

The symmetries (including $F^{\left(  0\right)  }$ itself) form
\emph{hierarchies} \cite{32}, \cite{33}. It is possible to establish
\emph{recursion relations} among the symmetries in each hierarchy. For many
equations (and all the equations our work deal with), the first symmetry is%

\begin{equation}
S^{\left(  1\right)  }\left[  u\right]  =u_{x} \label{eq 8}%
\end{equation}

Substituting the NIT (\ref{eq 2}) and the NF (\ref{eq 3}) into (\ref{eq 1})
leads to a sequence of \emph{homological equations} for the time evolution of
$u^{\left(  n\right)  }$, which have to be solved order-by-order.

The motivation for assumption (3) is that a perturbative analysis that does
not include the removal of resonant terms from the homological equation into
the NF, usually yields secularities, that is, unbounded terms in the
approximate solution. On the other hand, the NF is expected also to be
integrable and to preserve the nature of the unperturbed solution. This
feature is closely related to another significant one: that the main effect of
adding the higher-order terms to the NF is the update of physically-valuable
parameters (usually the wave velocity/the dispersion relation).

After removing the resonant terms out of the homological equations, they become:%

\begin{equation}
\left[  F^{\left(  0\right)  },\;u^{\left(  k\right)  }\left[  u\right]
\right]  +T^{\left(  k\right)  }\left[  u\right]  =0 \label{eq 6}%
\end{equation}
where $T^{\left(  k\right)  }\left[  u\right]  $ is the contribution for all
nonresonant terms of order $k$. The NIT is constructed from solutions of these equations.

However, the analysis may lead to the emergence of obstacles to integrability
\cite{15}-\cite{25}, beginning at some order in the expansion. These are terms
(differential polynomials) that the perturbative expansion of the dynamic
equation (Eq. \ref{eq 1}) generates, which cannot be accounted for by the
formalism. The differential polynomial structure of the obstacles is not
unique and depends on the way in which the NIT is constructed.

To make the construction of the NIT possible, the usual practice has been to
include these unaccounted-for terms in the NF. This makes the NF
nonintegrable, hence the name ``obstacles to integrability''. Including the
obstacles in the NF, disturbs the wave character of its solutions. The effect
of the obstacles in the case of the two-soliton solution of the normal form of
the perturbed KdV equation \cite{15}-\cite{17} has been studied in \cite{25}.
The zero-order solution was found to develop inelastic effects: appearance of
a second-order radiation wave; fourth-order, time dependent, corrections to
each of the wave numbers and the generation of an eighth-order soliton.

\bigskip

\section{Approach for overcoming obstacles}

\subsection{The general ideas}

The necessity to include the obstacles in the normal form is a consequence of
the assumption, usually made in the NF expansion, that all the terms in the
NIT are differential polynomials in the zero-order approximation (that is, in
$u\left(  x,t\right)  $) and do not depend explicitly on the independent
variables, $t$ and $x$. Our approach overcomes this problem by allowing the
higher-order terms in the NIT to depend on these variables. To this end, we
assume for the \textit{k}'th order term in the NIT the following form:%

\begin{equation}
u^{\left(  k\right)  }=u_{d}^{\left(  k\right)  }\left[  u\right]
+u_{r}^{\left(  k\right)  }\left(  x,t\right)  \label{eq 7}%
\end{equation}
In Eq. (\ref{eq 7}), $u_{d}^{\left(  k\right)  }\left[  u\right]  $ is a
differential polynomial in $u$, and $u_{r}^{\left(  k\right)  }\left(
x,t\right)  $ depends explicitly on $x$ and $t$, and is expected to account
for the obstacles. Thus, substituting the assumption (\ref{eq 7}) in the
homological equation (\ref{eq 6}), we obtain:%

\begin{equation}
\left[  F^{\left(  0\right)  },\;u_{r}^{\left(  k\right)  }\left(  x,t\right)
\right]  +R^{\left(  k\right)  }\left[  u\right]  =0 \label{eq 13}%
\end{equation}
where $R^{\left(  k\right)  }\left[  u\right]  \;$stands for the obstacle of
order $k$.

Owing to the freedom inherent in the perturbative expansion, the construction
of $u_{d}^{\left(  k\right)  }\left[  u\right]  $ is not unique. Unless
$u_{d}^{\left(  k\right)  }\left[  u\right]  $ is chosen in an appropriate
manner, the resulting obstacle may not reflect the following features of
physical interest:

(i) Obstacles do not emerge in the case of single-wave solutions of the NF
\cite{25};

(ii) The expectation that obstacles emerge owing to interaction among waves in
the multi-wave case \cite{20}.

Both features are realized if $u_{d}^{\left(  k\right)  }\left[  u\right]  $
is chosen to have the structure of the differential polynomial that solves the
problem in the case of a single-wave solution of the NF. The choice proposed
for $u_{d}^{\left(  k\right)  }\left[  u\right]  $ leads to obstacles in a
''canonical'' form, expressed in terms of symmetries of the unperturbed
equation. The obstacles now vanish if one substitutes for $u$ the single-wave
solution of the NF. More important, as a result, they are expected to vanish
away from regions of wave interaction in the multi-wave case. The reason is
that away from the interaction regions, multi-wave solutions asymptote into a
sum of well-separated single-wave solutions.

\subsection{Construction of canonical obstacles}

Starting with $u_{d}^{\left(  k\right)  }\left[  u\right]  \;$that solves the
homological equation (\ref{eq 6}) for the single-wave, we find that the
canonical obstacles can be written in the following form%

\begin{equation}
R^{\left(  n\right)  }\left[  u\right]  =\sum_{\substack{k=3\\i+j=k}%
}^{n+g}\gamma_{k}^{n}\,\,f_{ij}\left[  u,\,\partial_{x}\right]  \,R_{ij}%
\left[  u\right]  \label{eq 19}%
\end{equation}%

\begin{equation}%
\begin{array}
[c]{c}%
R_{pq}\left[  u\right]  =S^{\left(  p\right)  }\left[  u\right]  \,G^{\left(
q\right)  }\left[  u\right]  -S^{\left(  q\right)  }\left[  u\right]
\,G^{\left(  p\right)  }\left[  u\right] \\
S^{\left(  n\right)  }\left[  u\right]  =\partial_{x}\,G^{\left(  n\right)
}\left[  u\right]
\end{array}
\label{eq 24}%
\end{equation}
where $\gamma_{k}^{n}\;$is a numerical coefficient, $\,f_{ij}\left[
u,\,\partial_{x}\right]  \;$- a differential operator of an appropriate weight
and $g\;$is the gap between the index of symmetry and the order of
perturbation. For example, if we define $F^{\left(  0\right)  }\;$as
$S^{\left(  2\right)  }\;$(this is a widely-accepted notation), then $g=2$.

The obstacles of Eq. (\ref{eq 19}) vanish identically for the case of
single-wave solutions of the NF. To see this, we exploit the fact that all the
symmetries are proportional to one another in the case of the
\emph{single-wave} solutions of the NF. For ``trivial'' boundary condition,
$u\left(  \xi\rightarrow-\infty\right)  =0$:%

\begin{equation}
S^{\left(  n\right)  }=\left(  -1\right)  ^{n+1}\,v_{0}^{n+1}\,S^{\left(
1\right)  } \label{eq 31}%
\end{equation}
This may be proven by induction for any hierarchy which is governed by a
linear recursion relation.

The proportionality of all the symmetries leads to a simple update of the
velocity of the solutions of the NF:%

\begin{equation}
v=\sum_{k\geq0}\epsilon^{k}v_{k}\;,\;\;\;\;v_{k}=\left(  -1\right)
^{k}\,a_{k}\,v_{0}^{k+1} \label{eq 10}%
\end{equation}
Eq. (\ref{eq 10}) also describes the velocity update of each wave in a
multi-wave solution.

\subsection{Resonant contribution in obstacles?}

For multi-wave solutions of the NF, the obstacles do not vanish. An important
characteristic of our canonical obstacles is that they do not vanish only in
the interaction regions in the $x-t$ plane. For example, in the case of KdV
solitons, the interaction region is a finite domain around the origin, whereas
in the case of Burgers fronts it consists of one ore more domains of finite
width along semi-infinite lines. The canonical obstacles vanish exponentially
fast away from the interaction regions, where the solution asymptotes to
well-separated single waves. On the other hand, non-canonical obstacles are
finite also outside the interaction region.

A cardinal question that now arises is whether the obstacles generate secular
terms in the NIT (\ref{eq 2}). A symmetry, if contained in an obstacle, will
generate a secular term through Eq. (\ref{eq 13}). We, therefore, propose to
break an obstacle into a sum of a symmetry plus a non-resonant term. This
symmetry, with its coefficient, must be included in the NF (\ref{eq 3}).

Our task is therefore to determine whether a canonical obstacle has the
capacity of generating secular terms. A simple criterion for detecting this
capacity is that the obstacle spreads over an infinite or a semi-infinite
domain, and asymptotes to a symmetry. (This criterion is similar, although
much less rigorous, than the Fredholm Alternative Theorem.)

We focus on two-wave solutions of two equations: the Korteweg-de Vries (KdV)
equation with a two-soliton solution, with obstacle appearing in the second
order; and the Burgers equation, with a two-front solution, where an obstacle
arises already in the first order.

The interaction region, and thus the obstacles, of KdV are localized. Hence,
we expect the solution of Eq. \ref{eq 13} for this problem to be bounded. This
expectation has been verified by solving the homological equation numerically.
We obtain a new bounded soliton-like wave.

On the other hand, the interaction region in the two-front solution of the
Burgers equation is semi-infinite (the fronts are well separated in one half
of the $x-t$ plane and merge (interacting) in the other half). The canonical
obstacle asymptotically approaches the symmetry $S^{\left(  3\right)  }\left[
u\right]  $. Therefore, we expect the Burgers obstacle to generate a secular
term in Eq (\ref{eq 13}). This expectation has been verified by numerically
solving the equation.

We ''extract'' the symmetry $S^{\left(  3\right)  }\left[  u\right]  $ out of
the obstacle with its coefficient, and transfer it into the NF (\ref{eq 3}).
Thus, the NF remains solvable, but the coefficient of the first-order term in
the wave-velocity update is changed. The remainder is not a canonical
obstacle, that is, it is not confined to the interaction region but rather
spreads over all the fronts of the zero-order solution. However, it does not
contain a symmetry. The solution of the homological equation must be bounded.
There is no closed form solution of the homological equation, but the
numerical calculations show that this prediction is verified.

In the following section, these two examples are brought in some detail. The
full results of our work will be brought in further publications. We stress
that our results crusually depend on the wave nature of the solutions.

\subsection{Summary and Conclusions}

\begin{enumerate}
\item The main effect of perturbations on the interaction among waves in
multi-wave solutions of NL EPDE's is the emergence of the obstacles to integrability.

\item The proper place for handling obstacles, once their resonant part has
been shifted to the NF, is the homological equation. To this end, it is
necessary to allow the higher-order terms in the NIT (\ref{eq 2}) to be
explicit functions of the independent variables. Actually, $u^{\left(
n\right)  }\;$will consist of two parts: the differential polynomial,
$u_{d}^{\left(  n\right)  }\left[  u\right]  $, whose structure corresponds to
the case of a single-wave solution of the NF (\ref{eq 3}) (in this case, there
are no obstacles); and the function $u_{r}^{\left(  n\right)  }\left(
x,t\right)  \;$that is supposed to account for the obstacle.

\item When the NIT is built in such a way, the obstacles (if they exist) are
obtained in the canonical form (\ref{eq 19}). Canonical obstacles are confined
to the region of interaction among waves.

\item When the interaction region is localized (e.g., for KdV solitons), the
canonical obstacles do not generate secular solutions of the homological
equations (\ref{eq 13}). The solution of the homological equation is then
bounded, usually unavailable in closed form.

\item When the interaction region spreads over a semi-infinite range, the
obstacles are expected to cause secular solutions. Then, it is necessary to
identify a symmetry ''hidden'' inside the obstacle and remove it from the
homological equation into the NF. The remainder of the obstacle remains in the
homological equation and yields bounded solutions.
\end{enumerate}

In both situations, the obstacles cease to be obstacles to integrability of
the NF. The latter remains integrable, and its solutions (the zero-order
approximation) retain the character of the unperturbed solutions. The
difference between the items 4 and 5 above is that in \# 5 the wave velocity
is also affected, in the order in which the obstacle exists. Focusing on
two-wave solutions, it is found that the obstacle usually yields an additional
wave, that may not be expressed in closed form. In general, the homological
equation has to be solved numerically.

\bigskip

\section{Two worked examples}

\subsection{The perturbed KdV equation}

The perturbed KdV equation is%

\begin{equation}%
\begin{array}
[c]{c}%
w_{t}=6ww_{x}+w_{xxx}+\epsilon\left(  30\alpha_{1}w^{2}w_{x}+10\alpha
_{2}ww_{xx}+20\alpha_{3}w_{x}w_{xx}+\alpha_{4}w_{5x}\right) \\
+\epsilon^{2}\left(
\begin{array}
[c]{c}%
140\beta_{1}w^{3}w_{x}+70\beta_{2}w^{2}w_{xxx}+280\beta_{3}ww_{x}%
w_{xx}+14\beta_{4}ww_{5x}\\
+70\beta_{5}w_{x}^{3}+42\beta_{6}w_{x}w_{4x}+70\beta_{7}w_{xx}w_{xxx}%
+\beta_{8}w_{7x}%
\end{array}
\right)  +O\left(  \epsilon^{3}\right)
\end{array}
\label{eq 21}%
\end{equation}

We assume the NIT%

\begin{equation}
w=u+\epsilon u^{\left(  1\right)  }+\epsilon^{2}u^{\left(  2\right)
}+O\left(  \epsilon^{3}\right)  \label{eq 22}%
\end{equation}
and the NF%

\begin{equation}
u_{t}=S^{\left(  2\right)  }\left[  u\right]  +\epsilon\alpha_{4}S^{\left(
3\right)  }\left[  u\right]  +\epsilon^{2}\beta_{8}S^{\left(  4\right)
}\left[  u\right]  +O\left(  \epsilon^{3}\right)  \label{eq 23}%
\end{equation}
where%

\begin{equation}%
\begin{array}
[c]{c}%
S^{\left(  2\right)  }\left[  u\right]  =6u\,u_{x}+u_{xxx}\\
S^{\left(  3\right)  }\left[  u\right]  =30u^{2}u_{x}+10u\,u_{xx}%
+20u_{x}u_{xx}+u_{5x}\\%
\begin{array}
[c]{c}%
S^{\left(  4\right)  }\left[  u\right]  =140u^{3}u_{x}+70u^{2}u_{xx}%
+280u_{x}u_{xx}+14u\,u_{5x}\\
+70u_{x}^{3}+42u_{x}u_{4x}+70u_{xx}u_{xxx}+u_{7x}%
\end{array}
\end{array}
\label{eq 15}%
\end{equation}

A single-wave solution of the NF (\ref{eq 23}) is the well-known KdV soliton:%

\begin{equation}
u\left(  x,t\right)  =\frac{2k^{2}}{\cosh^{2}\left[  k\left(  x-vt+x_{0}%
\right)  \right]  } \label{eq 17}%
\end{equation}
and the two-wave solution is given by the Hirota formula \cite{28}:%

\begin{equation}%
\begin{array}
[c]{c}%
u\left(  x,t\right)  =2\partial_{x}^{2}\;\ln\left\{  1+g_{1}\left(
x,t\right)  +g_{2}\left(  x,t\right)  +\left(  \frac{k_{1}-k_{2}}{k_{1}+k_{2}%
}\right)  g_{1}\left(  x,t\right)  g_{2}\left(  x,t\right)  \right\} \\
\left(  g_{i}\left(  x,t\right)  =\exp\left[  2k_{i}\left(  x-v_{i}%
t+x_{0,i}\right)  \right]  \right)
\end{array}
\label{eq 18}%
\end{equation}
A sample of this solution is shown in Fig. 1. The only difference between the
solutions of the unperturbed KdV equation, $w_{t}=6ww_{x}+w_{xxx},\;$and the
both solutions of the NF is the update of the wave velocity:%

\begin{equation}
v_{i}=-4k_{i}^{2}-16\epsilon\alpha_{4}k_{i}^{4}-64\epsilon^{2}\beta_{8}%
k_{i}^{6}-O\left(  \epsilon^{3}\right)  \label{eq 20}%
\end{equation}

There are no obstacles in the first order in the Normal Form analysis.
However, an obstacle appears at the second order. Choosing the second-order
term in the NIT to be%

\begin{equation}%
\begin{array}
[c]{c}%
u^{\left(  2\right)  }=u_{r}^{\left(  2\right)  }\left(  x,t\right)
+B_{1}u^{3}+B_{2}u\,u_{xx}+B_{3}u_{x}^{2}+B_{4}u_{xxxx}\\
+B_{5}u\,u_{x}q^{\left(  1\right)  }+B_{6}u_{xxx}q^{\left(  1\right)  }%
+B_{7}u_{xx}q^{\left(  1\right)  ^{2}}+B_{8}u_{x}q^{\left(  2\right)  }\\
\left(  q^{\left(  1\right)  }\equiv\partial_{x}^{-1}u,\;\;\;\;\;\;q^{\left(
2\right)  }\equiv\partial_{x}^{-1}\left(  u^{2}\right)  \right)
\end{array}
\label{eq 25}%
\end{equation}
with the appropriate set of values for $\left\{  B_{k}\right\}  \;$based on
the form of $u^{\left(  2\right)  }$ in the case of the single-soliton
solution, we obtain the canonical obstacle%

\begin{equation}
R^{\left(  2\right)  }=\gamma_{3}^{2}uR_{21}=\gamma_{3}^{2}u\left(
3u^{2}u_{x}+u\,u_{xxx}-u_{x}u_{xx}\right)  \label{eq 26}%
\end{equation}
($\gamma_{3}^{2}\;$is built of a combination of coefficients of Eq. (\ref{eq
21}). This obstacle is localized (see Fig. 2) and hence cannot generate a
secular solution in the homological equation%

\begin{equation}
\partial_{t}u_{r}^{\left(  2\right)  }=6\partial_{x}\left(  u\,u_{r}^{\left(
2\right)  }\right)  +\partial_{x}^{3}u_{r}^{\left(  2\right)  }+\gamma_{3}%
^{2}uR_{21} \label{eq 27}%
\end{equation}
This equation can be solved in closed form by the Green's function method
developed in the context of the Inverse Scattering approach \cite{2},
\cite{35}-\cite{37}. As our goal is only to show that the solution of Eq.
(\ref{eq 27}), a numerical solution was sufficient for our purpose. It shows a
new bounded soliton-like wave (Fig. 3).

\subsection{The perturbed Burgers equation}

The perturbed Burgers equation is given by%

\begin{equation}
w_{t}=2ww_{x}+w_{xx}+\epsilon\left(  3\alpha_{1}w^{2}w_{x}+3\alpha_{2}%
ww_{xx}+3\alpha_{3}w_{x}^{2}+\alpha_{4}w_{xxx}\right)  +O\left(  \epsilon
^{2}\right)  \label{eq 28}%
\end{equation}
(we don't write here the second-order perturbation because an obstacle emerges
already in the first order). We assume, again, the NIT%

\begin{equation}
w=u+\epsilon u^{\left(  1\right)  }+O\left(  \epsilon^{2}\right)
\label{eq 29}%
\end{equation}
and the NF%

\begin{equation}
u_{t}=2u\,u_{x}+u_{xx}+\epsilon\mu\left(  3u^{2}u_{x}+3u\,u_{xx}+3u_{x}%
^{2}+u_{xxx}\right)  +O\left(  \epsilon^{2}\right)  \label{eq 30}%
\end{equation}

Its single-wave solution is the shock front%

\begin{equation}
u\left(  x,t\right)  =\frac{kA\exp\left[  k\left(  x-vt\right)  \right]
}{1+kA\exp\left[  k\left(  x-vt\right)  \right]  } \label{eq 11}%
\end{equation}
and the two-front solution is a straightforward extention of Eq. (\ref{eq 11}):%

\begin{equation}
u\left(  x,t\right)  =\frac{k_{1}A_{1}\exp\left[  k_{1}\left(  x-v_{1}%
t\right)  \right]  +k_{2}A_{2}\exp\left[  k_{2}\left(  x-v_{2}t\right)
\right]  }{1+k_{1}A_{1}\exp\left[  k_{1}\left(  x-v_{1}t\right)  \right]
+k_{2}A_{2}\exp\left[  k_{2}\left(  x-v_{2}t\right)  \right]  } \label{eq 12}%
\end{equation}
\bigskip(its sample is shown in Fig. 4, and one can see that the fronts are
interacting (i.e. merged) over a semi-infinite region). The velocity update
is, again, similar for the both solutions:%

\begin{equation}
v_{i}=-k_{i}-\epsilon\mu k_{i}^{2}-O\left(  \epsilon^{2}\right)  \label{eq 9}%
\end{equation}

The obstacle appears in the first-order analysis. If we move \emph{all} the
linear term $u_{xxx}\;$from the first order into the NF, that is, take
$\mu=\alpha_{4}$, and also choose $u^{\left(  1\right)  }\;$in the form that
solves the homological equation for the single-front case plus an explicit
function of $x\;$and $t$:%

\begin{equation}%
\begin{array}
[c]{c}%
u^{\left(  1\right)  }=u_{r}^{\left(  1\right)  }\left(  x,t\right)  +\left(
\alpha_{1}-2\alpha_{2}-\alpha_{3}+2\alpha_{4}\right)  qu_{x}-\frac{1}%
{2}\left(  2\alpha_{1}-\alpha_{2}+\alpha_{3}-2\alpha_{4}\right)  u^{2}\\
\left(  q\equiv\partial_{x}^{-1}u\right)
\end{array}
\label{eq 14}%
\end{equation}
then the obstacle will be get its canonical form:%

\begin{equation}%
\begin{array}
[c]{c}%
R^{\left(  1\right)  }=\gamma_{3}^{1}R_{21}=\gamma_{3}^{1}\left(  S^{\left(
2\right)  }G^{\left(  1\right)  }-S^{\left(  1\right)  }G^{\left(  2\right)
}\right)  =\gamma_{3}^{1}\left(  u\,S^{\left(  2\right)  }-u_{x}G^{\left(
2\right)  }\right) \\
\left(  \gamma_{3}^{1}=2\alpha_{1}-\alpha_{2}-2\alpha_{3}+\alpha_{4}\right)
\end{array}
\label{eq 41}%
\end{equation}
This obstacle is shown in Fig. 5, and one can see that it is finite over all
the interaction region. The homologicals equation now reads%

\begin{equation}
\partial_{t}u_{r}^{\left(  1\right)  }=2\partial_{x}\left(  u\,u_{r}^{\left(
1\right)  }\right)  +\partial_{x}^{2}u_{r}^{\left(  1\right)  }+\gamma_{3}%
^{1}R_{21} \label{eq 44}%
\end{equation}

It should be remarked that of the two terms of $R_{21}$, only $u\,S^{\left(
2\right)  }\;$may not be accounted for by the differential polyomials in the
NIT. On the contrary, it is easy to see that substituting $u_{r}^{\left(
1\right)  }=G^{\left(  2\right)  }\;$into the homogeneous part of the
homological equation yields%

\begin{equation}
-\partial_{t}G^{\left(  2\right)  }+2\partial_{x}\left(  u\,G^{\left(
2\right)  }\right)  +\partial_{x}^{2}G^{\left(  2\right)  }=2u_{x}G^{\left(
2\right)  } \label{eq 43}%
\end{equation}

As expected, the numerical solution of the homological equation shows the
existence of a secular term, which indicates the presence of a symmetry inside
$R_{21}$.

In order to extract this symmetry, we use the recursion relation among
symmetries in the Burgers hierarchy:%

\begin{equation}
S^{\left(  n+1\right)  }=S_{x}^{\left(  n\right)  }+u\,S^{\left(  n\right)
}+u_{x}G^{\left(  n\right)  }=S_{x}^{\left(  n\right)  }+R_{n1}+2u_{x}%
G^{\left(  n\right)  } \label{eq 33}%
\end{equation}
In particular, for $n=2$, it reads%

\begin{equation}
S^{\left(  3\right)  }=S_{x}^{\left(  2\right)  }+R_{21}+2u_{x}G^{\left(
2\right)  }\;\;\Rightarrow\;\;R_{21}=S^{\left(  3\right)  }-S_{x}^{\left(
2\right)  }-2u_{x}G^{\left(  2\right)  } \label{eq 35}%
\end{equation}
and the homological equation (\ref{eq 44}) becomes%

\begin{equation}
\partial_{t}u_{r}^{\left(  1\right)  }=2\partial_{x}\left(  u\,u_{r}^{\left(
1\right)  }\right)  +\partial_{x}^{2}u_{r}^{\left(  1\right)  }+\gamma_{3}%
^{1}\left(  S^{\left(  3\right)  }-S_{x}^{\left(  2\right)  }-2u_{x}G^{\left(
2\right)  }\right)  \label{eq 42}%
\end{equation}

Now, we choose $\mu$, the coefficient of $S^{\left(  3\right)  }$ in the NF,
to be
\begin{equation}
\mu=\alpha_{4}+\gamma_{3}^{1}=2\alpha_{1}-\alpha_{2}-2\alpha_{3}+2\alpha_{4}
\label{eq 40}%
\end{equation}
hence correcting the update of the wave velocity (Eq. \ref{eq 9}). Further, we
add an appropriate correction to $u_{d}^{\left(  1\right)  }\;$in order to
account for the term $-2\gamma_{3}^{1}u_{x}G^{\left(  2\right)  }\;$in Eq.
(\ref{eq 42}), according to Eq. (\ref{eq 43}).

Now, the equation for $u_{r}^{\left(  1\right)  }\;$becomes%

\begin{equation}
\partial_{t}u_{r}^{\left(  1\right)  }=2\partial_{x}\left(  u\,u_{r}^{\left(
1\right)  }\right)  +\partial_{x}^{2}u_{r}^{\left(  1\right)  }-\gamma_{3}%
^{1}S_{x}^{\left(  2\right)  } \label{eq 39}%
\end{equation}

The term $S_{x}^{\left(  2\right)  }\;$does not asymptote to any symmetry. The
numerical solution of Eq. (\ref{eq 39}) shows that a new bounded wave appears
(see Fig. 6).

\section{Acknowledgements}

G. Burde, L. Kalyakin and Y. Kodama are acknowledged.for helpful discussions.

\section{Figures}

\textbf{Figure captions}:

Fig. 1:\qquad Two-soliton solution of the KdV NF (Eq. \ref{eq 18});
$k_{1}=0.3,\;k_{2}=0.4.$

Fig. 2:\qquad Canonical obstacle $uR_{21}\;$for the KdV equation (Eq.\ref{eq
26}) for two-soliton solution; parameters as in Fig. 1.

Fig. 3:\qquad Contribution of this canonical obstacle to $u_{r}^{\left(
2\right)  }$ of Eq. (\ref{eq 27}) for zero boundary condition for
$x\rightarrow-\infty$; $k_{1}=0.5,\;k_{2}=0.75.$

Fig. 4:\qquad Two-front solution of the Burgers NF (Eq. \ref{eq 12});
$k_{1}=2,\;k_{2}=-2.$

Fig. 5:\qquad Canonical obstacle $-R_{21}\;$(with opposite sign) for the
Burgers equation (Eq. \ref{eq 41}) for the two-front solution; parameters as
in Fig. 4.

Fig. 6:\qquad Bounded contribution of the obstacle $S_{x}^{\left(  2\right)
}\;$to $u_{r}^{\left(  1\right)  }\;$of Eq. (\ref{eq 39}); parameters as in
Fig. 4.%

%TCIMACRO{\FRAME{itbpFU}{2.2355in}{1.7651in}{0in}{\Qcb{Fig. 1}}{}%
%{fig1.ps}{\special{ language "Scientific Word";  type "GRAPHIC";
%width 2.2355in;  height 1.7651in;  depth 0in;  original-width 6.5259in;
%original-height 5.1448in;  cropleft "0";  croptop "1";  cropright "1";
%cropbottom "0";  filename 'Fig1.eps';file-properties "NPEU";}}}%
%BeginExpansion
{\parbox[b]{2.2355in}{\begin{center}
\includegraphics[
height=1.7651in,
width=2.2355in
]%
{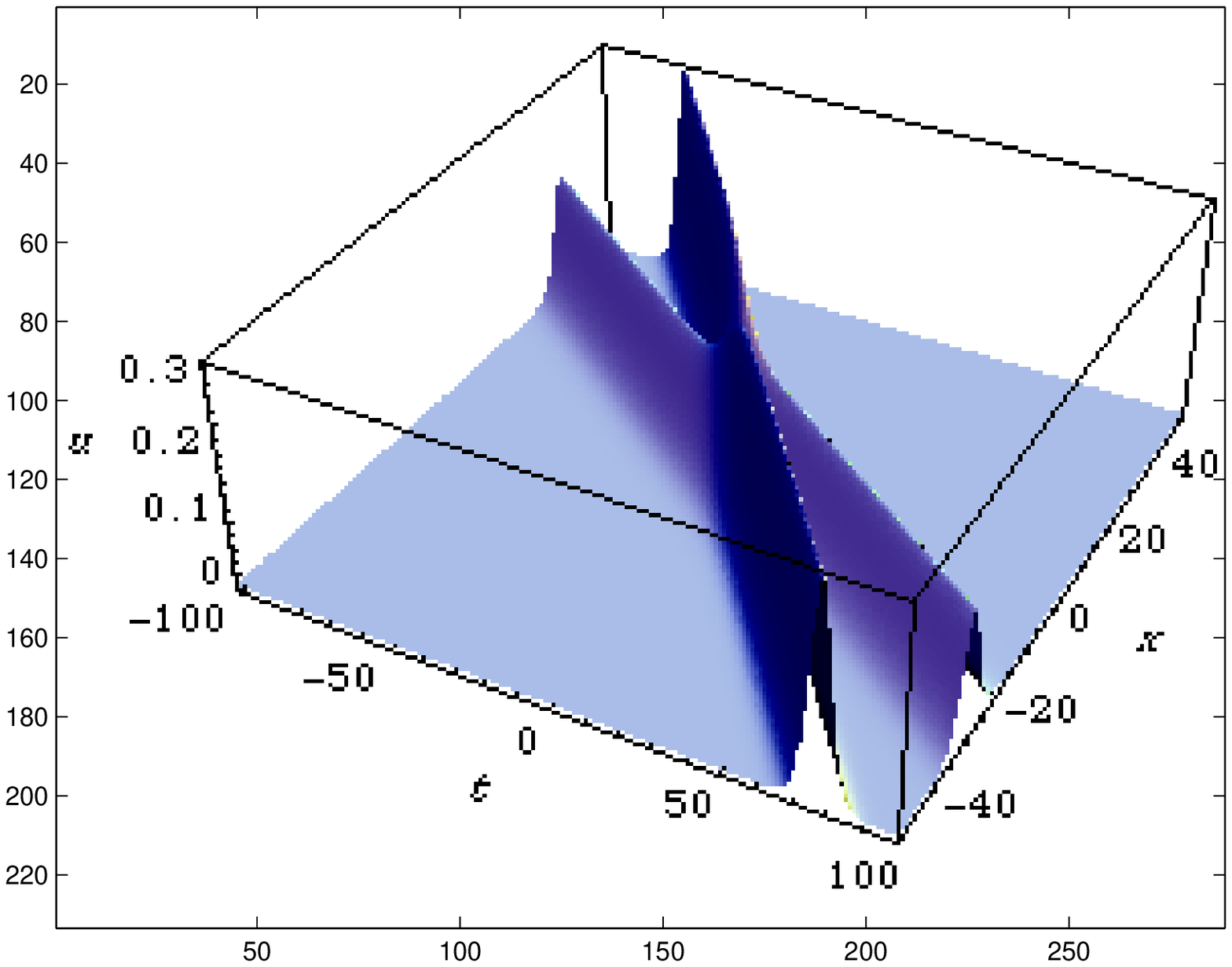}%
\\
Fig. 1
\end{center}}}%
%EndExpansion
\qquad%
%TCIMACRO{\FRAME{itbpFU}{2.2252in}{1.7564in}{0in}{\Qcb{Fig. 2}}{}%
%{fig2.ps}{\special{ language "Scientific Word";  type "GRAPHIC";
%width 2.2252in;  height 1.7564in;  depth 0in;  original-width 6.5259in;
%original-height 5.1448in;  cropleft "0";  croptop "1";  cropright "1";
%cropbottom "0";  filename 'Fig2.eps';file-properties "NPEU";}}}%
%BeginExpansion
{\parbox[b]{2.2252in}{\begin{center}
\includegraphics[
height=1.7564in,
width=2.2252in
]%
{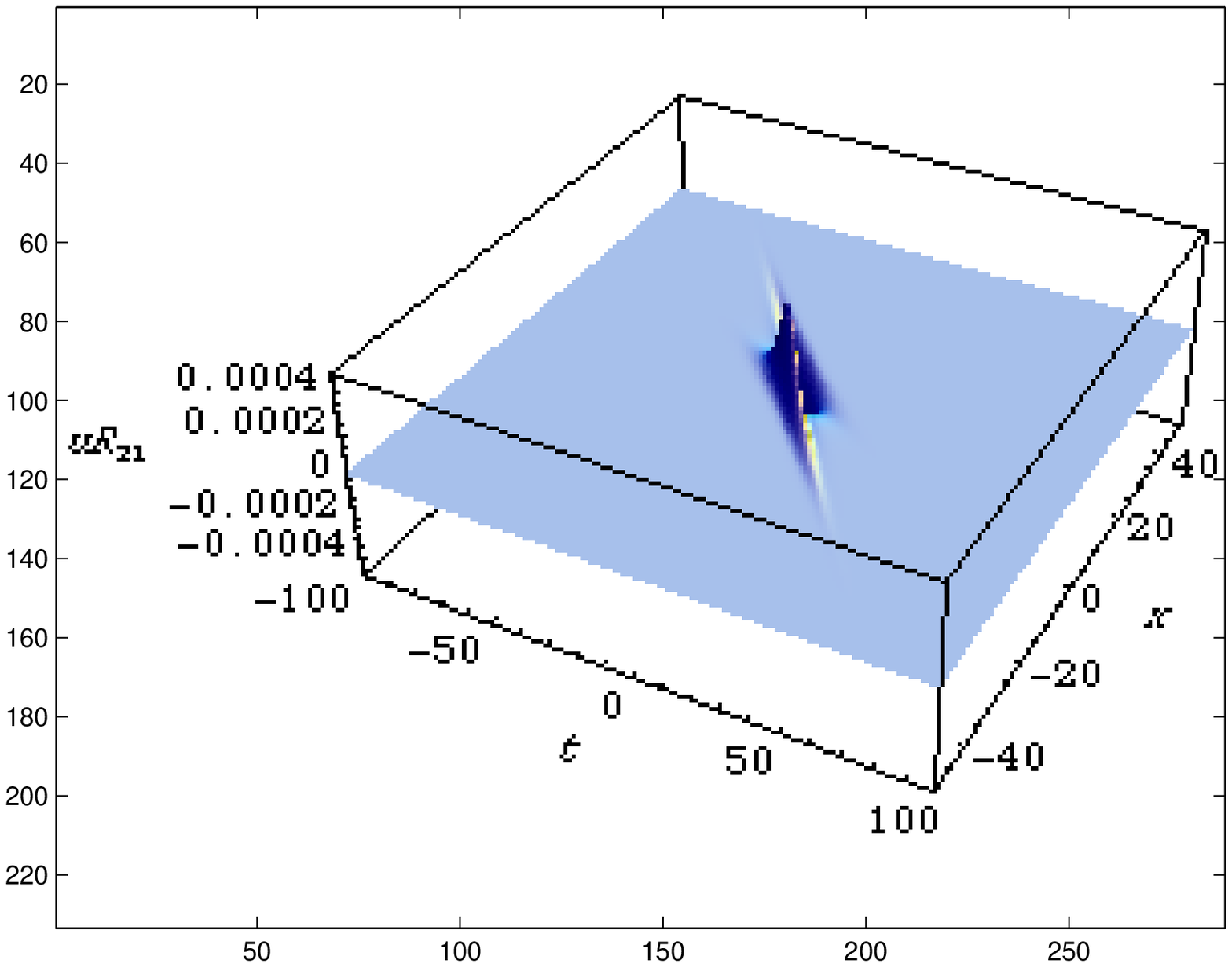}%
\\
Fig. 2
\end{center}}}%
%EndExpansion
\ \ \ \ \ \ 

\ \ \ %

%TCIMACRO{\FRAME{itbpFU}{2.2572in}{1.7807in}{0in}{\Qcb{Fig. 3}}{}%
%{fig3.ps}{\special{ language "Scientific Word";  type "GRAPHIC";
%width 2.2572in;  height 1.7807in;  depth 0in;  original-width 6.5259in;
%original-height 5.1448in;  cropleft "0";  croptop "1";  cropright "1";
%cropbottom "0";  filename 'Fig3.eps';file-properties "NPEU";}}}%
%BeginExpansion
{\parbox[b]{2.2572in}{\begin{center}
\includegraphics[
height=1.7807in,
width=2.2572in
]%
{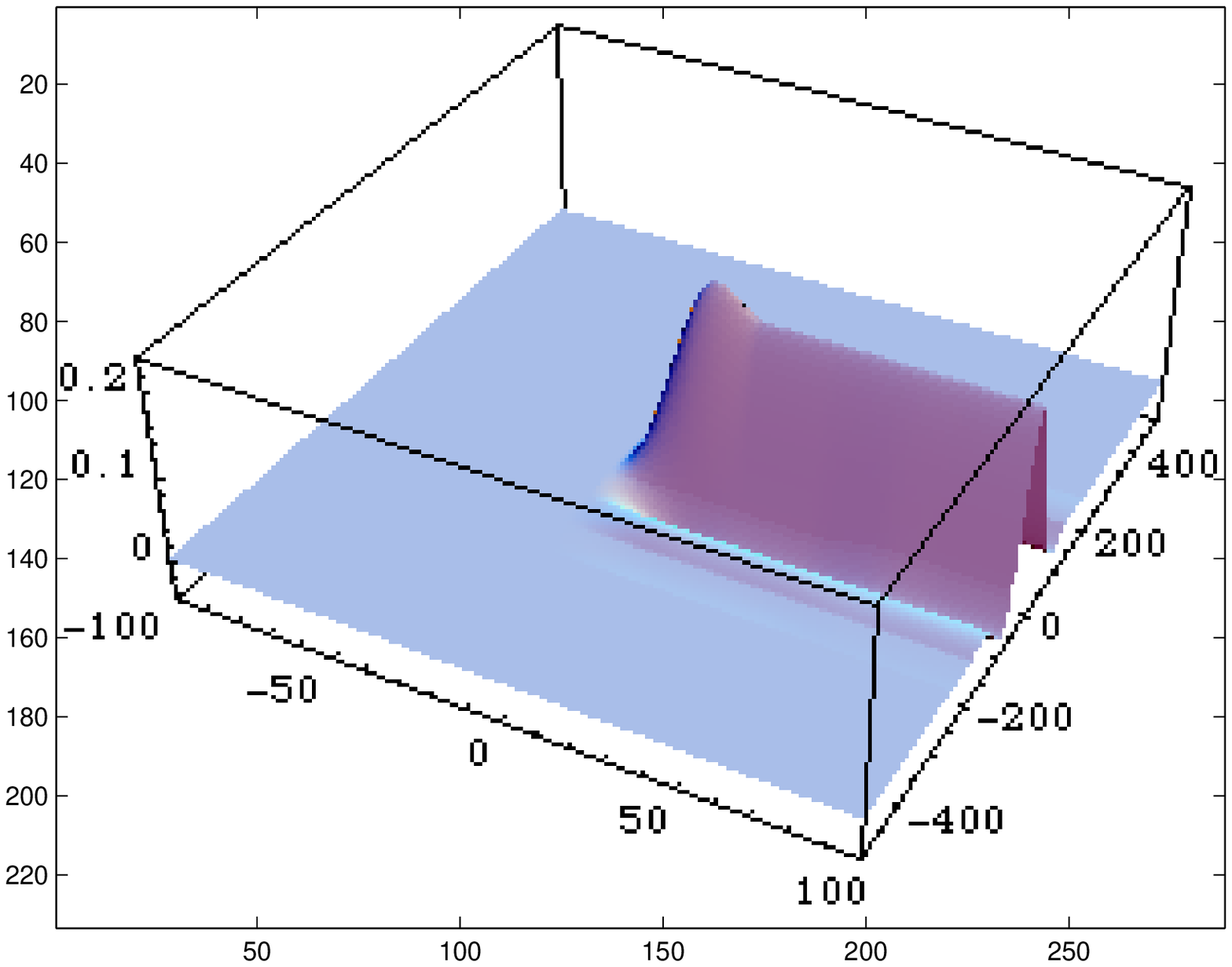}%
\\
Fig. 3
\end{center}}}%
%EndExpansion
\qquad%
%TCIMACRO{\FRAME{itbpFU}{2.2451in}{1.7729in}{0in}{\Qcb{Fig. 4}}{}%
%{fig4.ps}{\special{ language "Scientific Word";  type "GRAPHIC";
%width 2.2451in;  height 1.7729in;  depth 0in;  original-width 6.5259in;
%original-height 5.1448in;  cropleft "0";  croptop "1";  cropright "1";
%cropbottom "0";  filename 'Fig4.eps';file-properties "NPEU";}}}%
%BeginExpansion
{\parbox[b]{2.2451in}{\begin{center}
\includegraphics[
height=1.7729in,
width=2.2451in
]%
{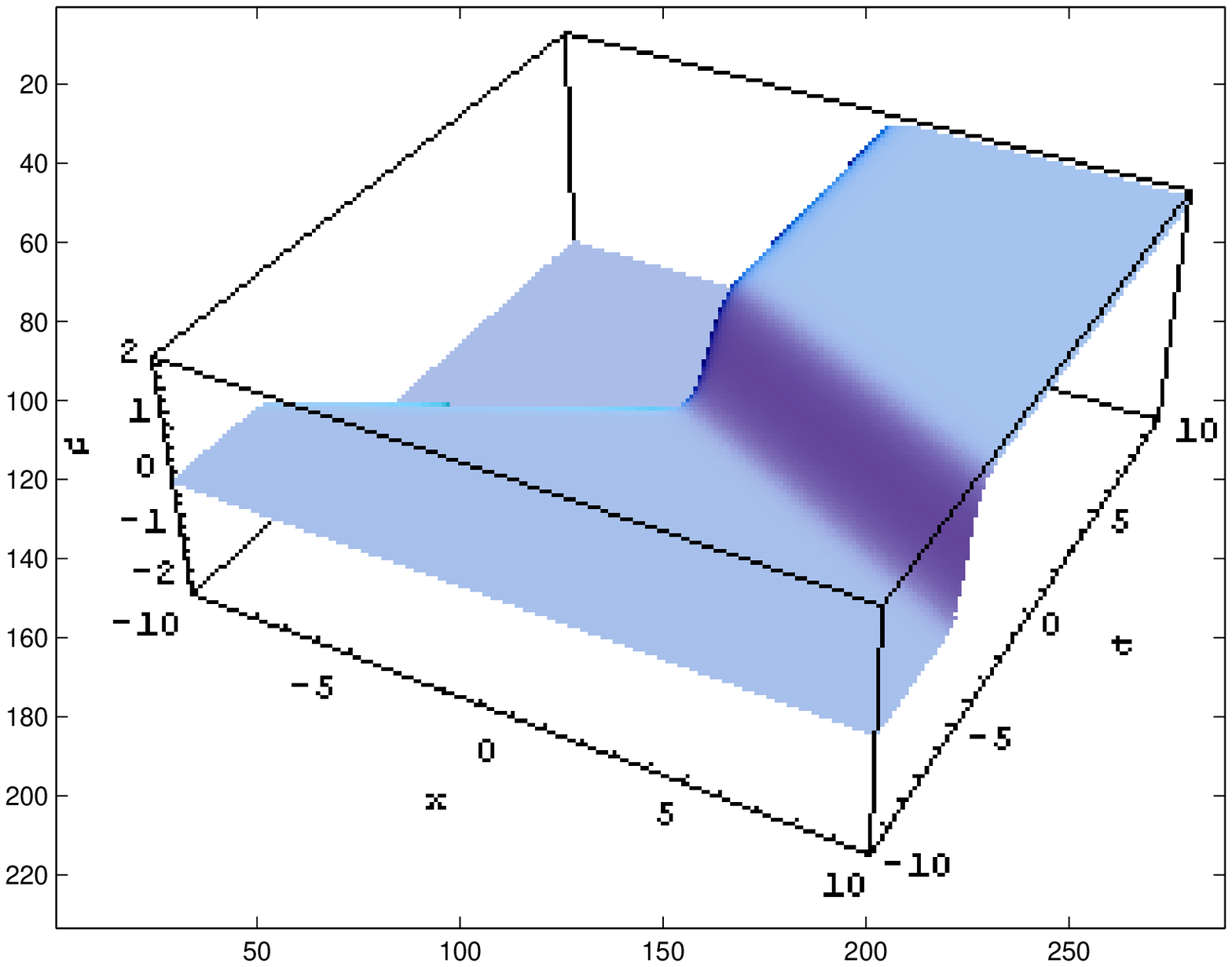}%
\\
Fig. 4
\end{center}}}%
%EndExpansion%

%TCIMACRO{\FRAME{itbpFU}{2.2779in}{1.7979in}{0in}{\Qcb{Fig. 5}}{}%
%{fig5.ps}{\special{ language "Scientific Word";  type "GRAPHIC";
%width 2.2779in;  height 1.7979in;  depth 0in;  original-width 6.5259in;
%original-height 5.1448in;  cropleft "0";  croptop "1";  cropright "1";
%cropbottom "0";  filename 'Fig5.eps';file-properties "NPEU";}}}%
%BeginExpansion
{\parbox[b]{2.2779in}{\begin{center}
\includegraphics[
height=1.7979in,
width=2.2779in
]%
{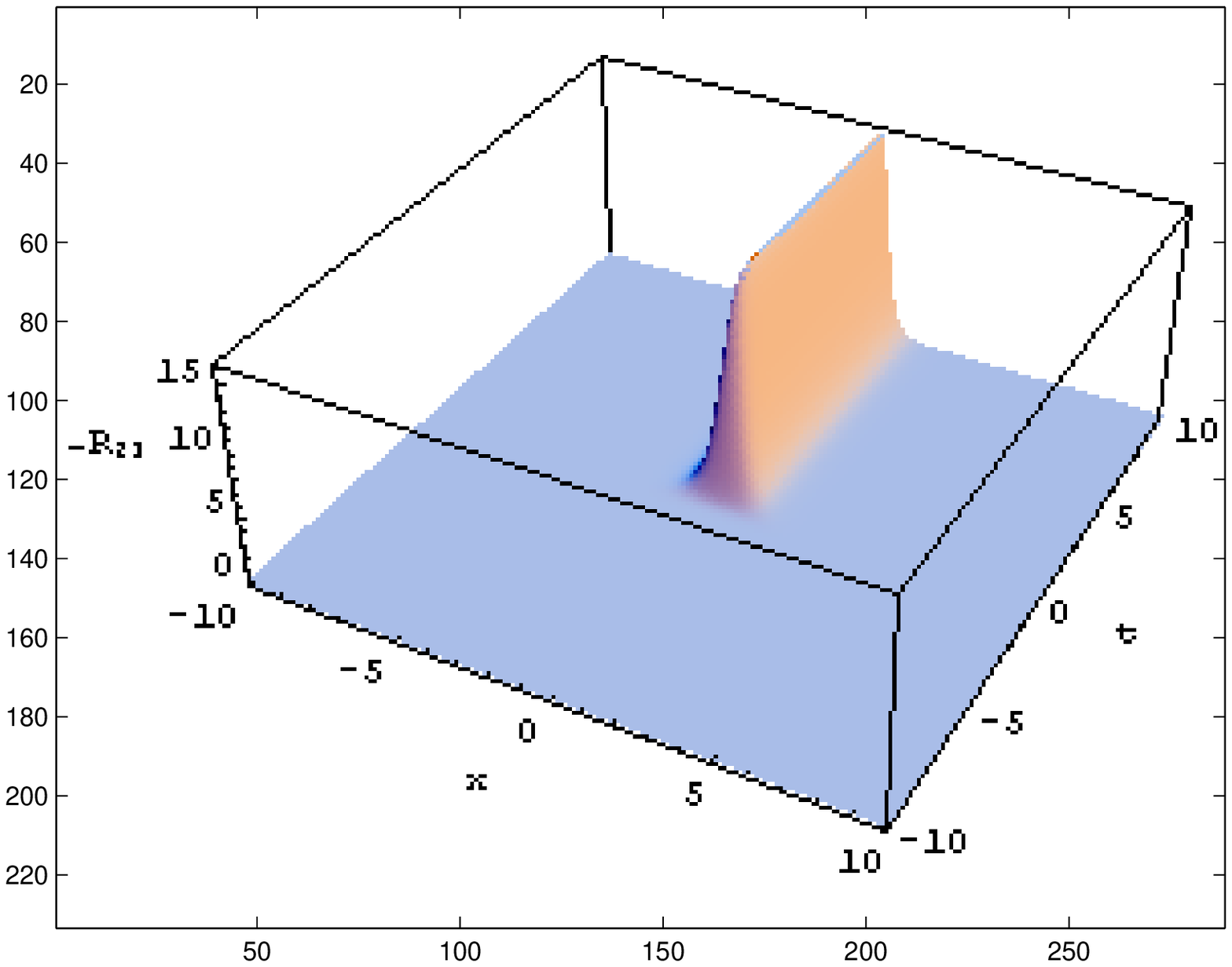}%
\\
Fig. 5
\end{center}}}%
%EndExpansion
\qquad%
%TCIMACRO{\FRAME{itbpFU}{2.2572in}{1.7807in}{0in}{\Qcb{Fig. 6}}{}%
%{fig6.ps}{\special{ language "Scientific Word";  type "GRAPHIC";
%width 2.2572in;  height 1.7807in;  depth 0in;  original-width 6.5259in;
%original-height 5.1448in;  cropleft "0";  croptop "1";  cropright "1";
%cropbottom "0";  filename 'Fig6.eps';file-properties "NPEU";}}}%
%BeginExpansion
{\parbox[b]{2.2572in}{\begin{center}
\includegraphics[
height=1.7807in,
width=2.2572in
]%
{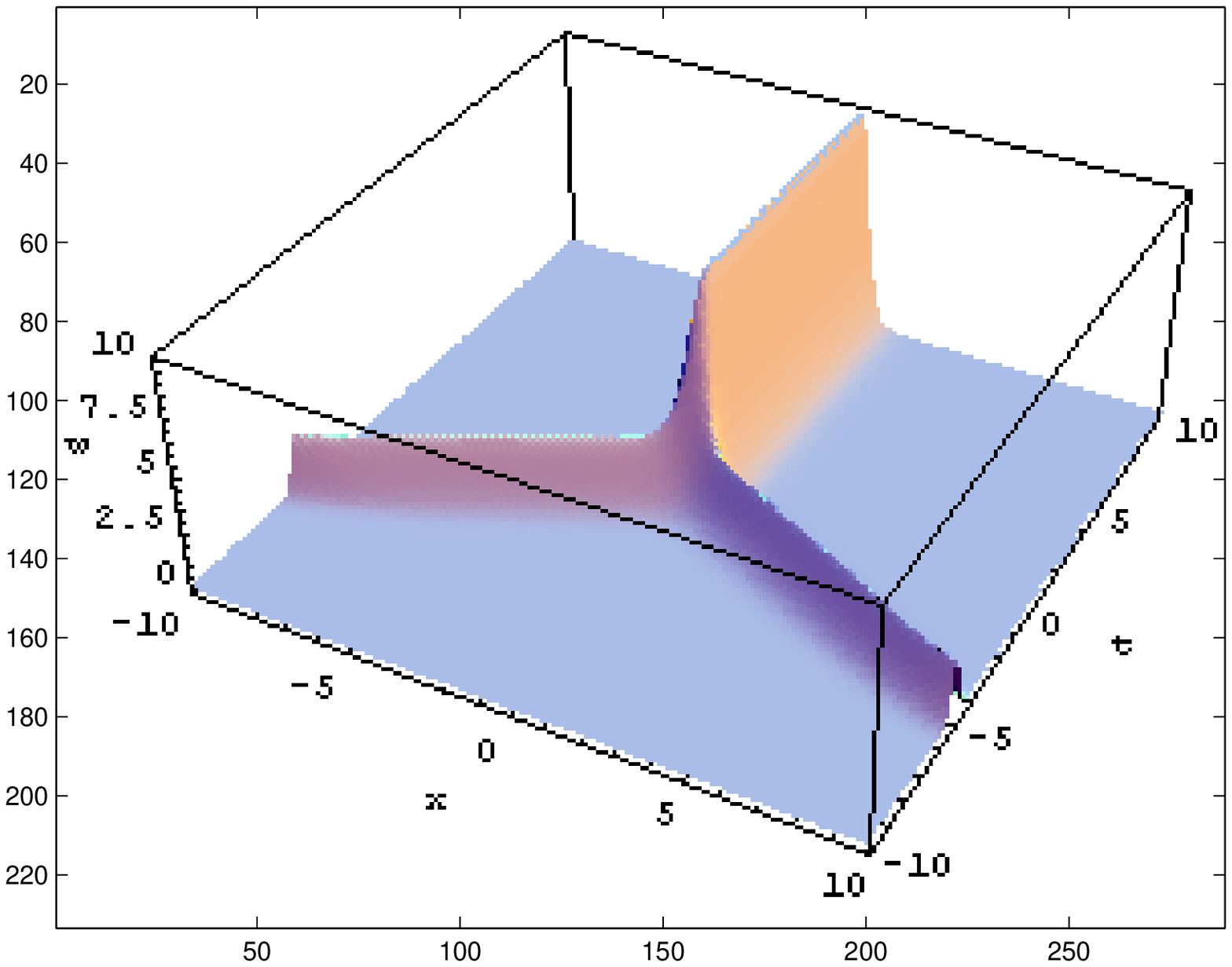}%
\\
Fig. 6
\end{center}}}%
%EndExpansion

\end{document}